\documentclass[twocolumn,showpacs,preprintnumbers,amsmath,amssymb]{revtex4}

\usepackage{graphicx}
\usepackage{dcolumn}
\usepackage{bm}

\begin{document}

\preprint{APS/123-QED}

\title{Effect of Long-Range Interactions
on the Multicritical Behavior of Homogeneous Systems}

\author{S.V. Belim}
 \email{belim@univer.omsk.su}
\affiliation{%
Omsk State University, 55-a, pr. Mira, Omsk, Russia, 644077
\textbackslash\textbackslash
}%

\date{\today}

\begin{abstract}
A field-theoretic approach is applied to describe behavior of
homogeneous three-dimensional systems with long-range interactions
defined by two order parameters at bicritical and tetracritical
points. Renormalization- group equations are analyzed in the
two-loop approximation by using the Pade–Borel summation
technique. The fixed points corresponding to various types of
multicritical behavior are determined. It is shown that effects
due to long-range interactions can be responsible for a change
from bicritical to tetracritical behavior.

\end{abstract}

\maketitle
It was shown in [1] that effects due to a long-range
interaction described by the power law $r^{ – D – \sigma}$ with $\sigma < 2$ are
responsible for a change in critical behavior. It was also
revealed that the system is described by mean-field critical
exponents.

This paper deals with effects of long-range interaction
on systems described by two fluctuating order
parameters. The phase diagrams of such systems can
contain bicritical and tetracritical points. In the former
case, a multicritical point is the point of intersection of
two second-order phase transition curves and one first order
phase transition curve. In the latter case, it is the
point of intersection of four second-order phase transition
curves. In the neighborhood of a multicritical
point, the system exhibits critical behavior characterized
by competing types of ordering. Whereas one critical
parameter is replaced by the other at a bicritical
point, the ordering types can coexist in a mixed phase
at a tetracritical point. Such systems can be described in
terms of two order parameters that are transformed
under different irreducible representations.

The model Hamiltonian is written as
\begin{eqnarray}\label{gam}
&&H_0=\frac 12\int d^Dq(\tau _1+q^\sigma)\Phi_q\Phi_{-q}\\
&&+\frac 12\int d^Dq(\tau _2+q^a)\Psi_q\Psi_{-q}\nonumber\\
&&+u_{01}\int d^D\{q_i\}(\Phi_{q1}\Phi_{q2})(\Phi_{q3}\Phi_{-q1-q2-q3})\nonumber\\
&&+u_{02}\int d^D\{q_i\}(\Psi_{q1}\Psi_{q2})(\Psi_{q3}\Psi_{-q1-q2-q3})\nonumber\\
&&+2u_{03}\int d^D\{q_i\}(\Phi_{q1}\Phi_{q2})(\Psi_{q3}\Psi_{-q1-q2-q3})\nonumber
\end{eqnarray}
where $\Phi$ and $\Psi$ are fluctuating order parameters,
$u_{01}$ and $u_{02}$ are positive constants,
$\tau_1\sim|T-T_{c1}|/T_{c1}$, $\tau_2\sim|T-T_{c2}|/T_{c2}$,
$T_{c1}$ and $T_{c2}$ are the corresponding phase-transition temperatures,
and $\sigma$ is the long-range parameter.

This Hamiltonian admits a wide diversity of multicritical
points. The conditions for tetracritical and
bicritical behavior are $u_3^2<u_1u_2$ and $u_3^2\geq u_1u_2$, respectively.

In the field-theoretic approach [3], the asymptotic
critical behavior and structure of phase diagrams in the
fluctuation region are governed by the Callan–Symanczyk
equation for the vertex parts of irreducible Green
functions. The $\beta$ and $\gamma$ functions contained in the
Callan–Symanczyk equation for renormalized interaction
vertexes $u_1, u_2, u_3$ are calculated by a standard
method based on the Feynman diagram technique and
a renormalization procedure [4]. As a result, the following
expressions for $\beta$ functions are obtained in the two-loop
approximation:
\begin{eqnarray}\label{8}
&&\beta _{u1}=(2\sigma-D)(-u_1+36J_0u_1^2+4J_0u_3^2\\
&&-1728\Big(2J_1-J_0^2-\frac29G\Big)u_1^3\nonumber\\
&&-192\Big(2J_1-J_0^2-\frac29G\Big)u_1u_3^2\nonumber\\
&&-64\Big(2J_1-J_0^2)u_3^3),\nonumber\\
&&\beta _{u2}=(2\sigma-D)(-u_2+36J_0u_2^2+4J_0u_3^2\nonumber\\
&&-1728\Big(2J_1-J_0^2-\frac29G\Big)u_2^3\nonumber\\
&&-192\Big(2J_1-J_0^2-\frac29G\Big)u_2u_3^2\nonumber\\
&&-64\Big(2J_1-J_0^2)u_3^3),\nonumber\\
&&\beta _{u3}=(2\sigma-D)(-u_3+16J_0u_3^2+12J_0u_1u_3+12J_0u_2u_3\nonumber\\
&&-320\Big(2J_1-J_0^2-\frac25G\Big)u_3^3-\nonumber\\
&&-288\Big(2J_1-J_0^2-\frac23G\Big)u_1^2u_3\nonumber\\
&&-288\Big(2J_1-J_0^2-\frac23G\Big)u_2^2u_3\nonumber\\
&&-576\Big(2J_1-J_0^2\Big)u_1u_3^2-\nonumber\\
&&-576\Big(2J_1-J_0^2\Big)u_2u_3^2),\nonumber
\end{eqnarray}
\begin{eqnarray}
J_1&=&\int \frac{d^Dq d^Dp}{(1+|\vec{q}|^\sigma)^2(1+|\vec{p}|^\sigma)
(1+|q^2+p^2+2\vec{p}\vec{q}|^{\sigma/2})},\nonumber\\
J_0&=&\int \frac{d^Dq}{(1+|\vec{q}|^\sigma)^2},\nonumber\\
G&=&-\frac{\partial}{\partial |\vec{k}|^\sigma}\int \frac{d^Dq
d^Dp}{(1+|q^2+k^2+2\vec{k}\vec{q}|^\sigma)(1+|\vec{p}|^\sigma)}\nonumber\\
&&\cdot\frac1{(1+|q^2+p^2+2\vec{p}\vec{q}|^{\sigma/2})}\nonumber
\end{eqnarray}
In terms of the new effective interaction vertices
\begin{equation}\label{vertex}
    v_1=u_1\cdot J_0, \ \ \ \ v_2=u_2\cdot J_0, \ \ \ \ v_3=u_3\cdot J_0.
\end{equation}
the $\beta$ functions are expressed as
\begin{eqnarray}\label{9}
&&\beta _{1}=(2\sigma-D)(-v_1+36v_1^2+4v_3^2\\
&&-1728\Big(2\widetilde{J_1}-1-\frac29\widetilde{G}\Big)u_1^3\nonumber\\
&&-192\Big(2\widetilde{J_1}-1-\frac29\widetilde{G}\Big)v_1v_3^2\nonumber\\
&-&64\Big(2\widetilde{J_1}-1)v_3^3),\nonumber\\
&&\beta _{2}=(2\sigma-D)(-v_2+36v_2^2+4v_3^2\nonumber\\
&&-1728\Big(2\widetilde{J_1}-1-\frac29\widetilde{G}\Big)v_2^3\nonumber\\
&&-192\Big(2\widetilde{J_1}-1-\frac29\widetilde{G}\Big)v_2v_3^2\nonumber\\
&&-64\Big(2\widetilde{J_1}-1)v_3^3),\nonumber\\
&&\beta _{u3}=(2\sigma-D)(-v_3+16v_3^2+12v_1v_3+12v_2v_3\nonumber\\
&&-320\Big(2\widetilde{J_1}-1-\frac25\widetilde{G}\Big)v_3^3-\nonumber\\
&&-288\Big(2\widetilde{J_1}-1-\frac23\widetilde{G}\Big)v_1^2v_3\nonumber\\
&&-288\Big(2\widetilde{J_1}-1-\frac23\widetilde{G}\Big)v_2^2v_3
-576\Big(2\widetilde{J_1}-1\Big)v_1v_3^2\nonumber\\
&&-576\Big(2\widetilde{J_1}-1\Big)v_2v_3^2),\nonumber\\
&&\widetilde{J_1}=J_1/J_0^2\ \ \ \
 \widetilde{G}=G/J_0^2.\nonumber
\end{eqnarray}
This redefinition is meaningful for $\sigma\leq D/2$. In this
case, $J_0$, $J_1$ and $G$ are divergent functions. Introducing
the cutoff parameter $\Lambda$, we obtain finite expressions for
the ratios $J_1/J_0^2$ and $G/J_0^2$ as $\Lambda\rightarrow\infty$.

The integrals are performed numerically. For $\sigma\leq D/2$,
a sequence of $J_1/J_0^2$ and $G/J_0^2$ corresponding to
various values of $\Lambda$ is calculated and extrapolated to
infinity.

It is well known that perturbation-theory series are
asymptotic and expressions (4) cannot be applied
directly since the interaction vertexes for order-parameter
fluctuations in the fluctuation region are too large.
For this reason, the required physical information was
extracted from these expressions by applying the Pade–Borel
method extended to the four-parameter case. The
appropriate direct and inverse Borel transforms have
the form
\begin{eqnarray}
&& f(v1,v2,v3)=\sum\limits_{i_1,i_2,i_3}
c_{i_1i_2i_3}v_1^{i_1}v_2^{i_2}v_3^{i_3}\nonumber\\
&&=\int\limits_{0}^{\infty}e^{-t}F(v_1t,v_2t,v_3t,z_1t,z_2t,w_1t,w_2t)dt,  \\
&& F(v1,v2,v3)=\sum\limits_{i_1,i_2,i_3}
\frac{\displaystyle c_{i_1,i_2,i_3}}
{\displaystyle(i_1+i_2+i_3)!}v_1^{i_1}v_2^{i_2}v_3^{i_3}.
\end{eqnarray}
To obtain an analytic continuation of the Borel transform
of a function, a series in an auxiliary variable è is
introduced:
\begin{equation}
\tilde{F}(v_1,v_2,v_3,\theta)
=\sum\limits_{k=0}^{\infty}\theta^k\sum\limits_{i_1,i_2,i_3}
\frac{\displaystyle c_{i_1i_2i_3}}{\displaystyle k!}v_1^{i_1}v_2^{i_2}v_3^{i_3}
\delta_{i_1+i_2+i_3,k}\  ,
\end{equation}
The Pade approximant [L/M] is applied to this series at
$\theta=1$. The [2/1] approximant is used to calculate the $\beta$
functions in the two-loop approximation. The critical
behavior is completely determined by the stable fixed
point $(v_1^*,v_2^*,v_3^*)$ satisfying the system of equations
\begin{equation}
\beta_{i}(v_1^*,v_2^*,v_3^*)=0\ \ \ \ \ \ \ \ \ \ \ \ \ \   (i=1,2,3).
\end{equation}
The requirement of stability of a fixed point reduces to
the condition that the eigenvalues $b_i$ of the matrix
\begin{equation}  \displaystyle
B_{i,j}=\frac{\partial\beta_i(v_1^*,v_2^*,v_3^*)}{\partial{v_j}}\ \ \ \ \ \ (i=1,2,3)
\end{equation}
lie in the right half-plane.

The resulting system of summed $\beta$ functions has a
wide diversity of fixed points lying in the physical
region of vertexes with $v_i\geq 0$. A complete analysis of
the fixed points corresponding to the behavior of only
one order parameter was presented in [1]. Here, the
simultaneous critical behavior of both order parameters
is considered. The stable fixed points and eigenvalues
of the stability matrix are listed in the table.

An analysis of the critical points and their stability
leads to certain conclusions. When $\sigma>1.6$, bicritical
behavior is observed $v_3^2\geq v_1v_2$), as in the case of a
short-range interaction. When $1.5<a\leq 1.6$, tetracritical
behavior is observed ($v_3^2<v_1v_2$).

Thus, effects due to long-range interaction are
responsible for a change from bicritical to tetracritical
behavior when $1.5<a\leq 1.6$.

\begin{table}
\begin{center}
\begin{tabular}{|c|c|c|c|c|c|c|} \hline
$\sigma$&$v_1^{*}$& $v_2^{*}$ & $v_3^{*}$ &$b_1$  & $b_2$ &$b_3$    \\
\hline
1.9&0.035842 &  0.035842&  0.039202 &  0.069 & 0.505 & 0.702   \\
1.8&0.033682 &  0.033682&  0.034575 &  0.090 & 0.571 & 0.753   \\
1.7&0.031287 &  0.031287&  0.031334 &  0.113 & 0.629 & 0.809   \\
1.6&0.027427 &  0.027427&  0.026699 &  0.157 & 0.738 & 0.919   \\
1.5&0.026514 &  0.026514&  0.025973 &  0.171 & 0.762 & 0.949   \\
\hline
\end{tabular} \end{center} \end{table}

The work is supported by Russian Foundation for Basic Research N
04-02-16002.

\newpage
\def\baselinestretch{1.0}

\end{document}